\documentclass[a4paper,11pt]{article}
\usepackage{pos}
\usepackage[numbers]{natbib}
\usepackage[outdir=./]{epstopdf}

\title{First Year of Stellar-Mass Black Hole Observations with the Imaging X-ray Polarimetry Explorer}

\ShortTitle{{\it IXPE} Observations of Stellar-Mass Black Holes}

\author*{Nicole Rodriguez Cavero}
\onbehalf{on behalf of the {\it IXPE} Science Collaboration}

\affiliation[]{Physics Department, McDonnell Center for the Space Sciences, and Center for Quantum Leaps, Washington University in St. Louis,\\ St. Louis, MO 63130, USA}
\emailAdd{n.rodriguez@wustl.edu}

\abstract{The Imaging X-ray Polarimetry Explorer ({\it IXPE}), launched on December 9, 2021, enables X-ray polarimetric observations with unprecedented sensitivity in the 2-8 keV energy range. X-ray polarization allows us to test accretion disk, corona, and emission models of stellar-mass black holes in X-ray binaries found predominantly in soft and hard states of accretion. Every state is dominated by a combination of thermal disk, coronal, or reflected emission—each type containing different information about the environment around the black hole that can be understood through polarization. In 2022, {\it IXPE} measured the polarization signatures of four stellar-mass black holes: Cygnus X-1, 4U 1630-47, Cygnus X-3, and LMC X-1. We report on the physical consequences of these first {\it IXPE} observations and the science driven by this new chapter in X-ray polarimetry.}

\FullConference{
Multifrequency Behaviour of High Energy Cosmic Sources XIV (MULTIF2023)\\
12-17 June 2023\\
Palermo, Italy\\}

\begin{document}
\maketitle

\section{Introduction}

Black hole X-ray binaries (BHXRBs) consist of a stellar-mass black hole and a closely orbiting donor star. The matter transferred from the donor star to the accretor gives rise to structures like the black hole accretion disk, corona, and radio jet.
To study these components, we turn to X-ray polarization.
Polarimetry offers a unique geometrical insight into BHXRBs from which we can rectify the degeneracies between the black hole mass, spin, and inclination, constrain the shape of their coronae, and predict the origin of soft X-ray seed photons.

In the standard Shakura-Sunyaev approach, the structure of the accretion flow is driven by the conversion of gravitational potential to heat via viscous stresses \cite{Shakura-Sunyaev1973}. The accreted matter from the companion star forms an optically thick disk around the black hole which emits at a quasi-blackbody temperature of T($r$) $\sim r^{-3/4}$ producing a geometrically thin disk so long as the luminosities are well below Eddington \cite{Done2007}.
Around the inner accretion flow we also find the corona: a structure of geometrically thick, optically thin $\sim 100$ keV plasma that radiates in the form of Comptonized, bremsstrahlung, and/or synchrotron emission \cite{Done2007, McClintock2006}. Up until recent, several corona geometries and locations were proposed but unconstrained. The most common ones include the cone-shaped coronae in the funnel regions around the black hole spin axis, the sandwich or wedge-shaped coronae extended in the plane of the disk \cite{Schnittman2010}, the lamp-post coronae consisting of a spherical or conical body of plasma located on the black hole rotation axis and at some height above and below the disk \cite{Zhang2022}, and the hot inner flow-truncated disk coronae in which the inner  disk region is replaced with a region of hot plasma possibly through evaporation of the disk \cite{Gilfanov2010}.

BHXRBs can be found in different states of accretion typically defined by either their luminosity, power-law photon index, or the absence/presence of a jet \cite{Fender2004}.
Here, we are interested in the soft and hard states of accretion. In the high/soft state (HSS), also known as the thermal-dominant state, the total emission is dominated by thermal radiation from the accretion disk that peaks at around $\sim 1$~keV. We typically employ a multi-temperature blackbody model to explain the majority of the observed soft X-ray energy spectra \cite{Mitsuda1984}.
In the low/hard state (LHS), most of the X-ray emission comes from Compton scattering in the coronal plasma peaking at $\sim 100$~keV. During this state, the BHXRB energy spectra is dominated by a power-law component in the $5$---$20$ keV energy range with a photon index $\propto E^{-\Gamma}$ of $1.5 < \Gamma < 2$ \cite{Done2007}. Additionally, during the LHS, the total emission is also comprised of reflected emission from photons Comptonized in the corona bouncing off the disk giving rise to fluorescent emission lines \cite{Tomsick2014A}.
The LHS is also associated with the presence of a steady jet detected at radio frequencies \cite{Fender2004}. 
In the case of Cygnus X-3, however, the hard state is associated with radio quiescence \cite{Szostek2008} and exhibits a hard X-ray spectra peaking at $\sim 20$~keV \cite{Zdziarski2004}.
Stellar-mass black holes can also be found in the steep power-law state (SPL)---characterized by competing thermal and power-law emission \cite{Remillard2006}.

The polarization signature of stellar-mass black holes will vary depending on the state in which they are observed since the relative contribution to the emission from different components of the black hole will affect the direction of polarization. Numerical general relativistic ray-tracing and radiation transport codes \cite{Schnittman2009, Schnittman2010, Krawczynski2012, MONK} use Chandrasekhar results for the polarization of emission from a fully ionized atmosphere \cite{Chandrasekhar1960} to predict the X-ray polarization of stellar-mass black holes. 
In the thermal state, the expected polarization of the thermal disk emission is 0.1\%, 0.7\%, and 2.3\% for $i=20^{\circ}$, $40^{\circ}$, and $60^{\circ}$, respectively \cite{Chandrasekhar1960}. In the cases where returning radiation is considered, the PD is even smaller owing to the effect of gravitational lensing \cite{Dovciak2008}.
Direct thermal emission is expected to be polarized parallel to the accretion disk while thermal emission that is lensed around the black hole exhibits a 90$^{\circ}$ rotation and is polarized in the direction of the black hole spin axis \cite{Schnittman2009}.
Polarization observations of sources in the HSS allow us to test the standard thin disk model of accretion as well as constrain black hole spin and inclination by analyzing the dependence of polarization with respect to energy \cite{Schnittman2009,Li2009}.
Since coronal emission is expected to be polarized perpendicular to the direction of the scattering surface \cite{Schnittman2010}, polarization measurements in the LHS can be used to determine the properties of the coronal plasma as well as the inner accretion flow

The Imaging X-ray Polarimetry Explorer ({\it IXPE}, \cite{Weisskopf2022}) is a collaboration between NASA and the Italian Space Agency. {\it IXPE} is the first X-ray polarization space observatory since the eight Orbiting Solar Observatory ({\it OSO-8} \cite{OSO-8}) and it is 30 times more sensitive for a 10 mCrab source \cite{Soffita2020}. The instrument is comprised of three gas pixel detector (GPD) units, each paired with a focusing grazing incidence mirror at a 4 meter focal length \cite{DiMarco2022}. The GPD provides a 1-2 $\mu$s timing accuracy and an angular resolution of $\leq 30$ arc-seconds \cite{Soffita2020}. {\it IXPE} grants simultaneous timing, imaging, spectral, and polarimetric measurements in the 2-8~keV energy band---making it a powerful tool to study astrophysical sources.
In 2022, {\it IXPE} observed four stellar-mass black holes located in X-ray binaries. The highly accurate polarization measurements of these observations have yielded impressive results thus far.
In the following, we discuss the four {\it IXPE} stellar-mass black hole observations that took place in 2022 and their implications on the structure of the black hole system and its emission mechanisms.

\section{Results}

The first year of {\it IXPE} observations of BHXRBs revealed higher-than-expected linear polarization degrees (PDs) that increased with energy and linear polarization angles (PAs) that remained relatively constant in the 2-8 keV energy band. These high precision measurements were accompanied by quasi-simultaneous observations from other X-ray instruments: most commonly NICER \cite{nicer} and {\it NuSTAR} \cite{nustar}.
Cygnus X-1 and Cygnus X-3 were observed in the LHS where black hole coronal and reflected emission tend to dominate. LMC X-1 and 4U 1630-47, on the other hand, were observed in the soft state where thermal emission from accretion disk dominates the signal.
We highlight the results of these observational campaigns.

\subsection{Cygnus X-1}

\begin{figure}[t!]
\begin{center}
\includegraphics[width=14cm]{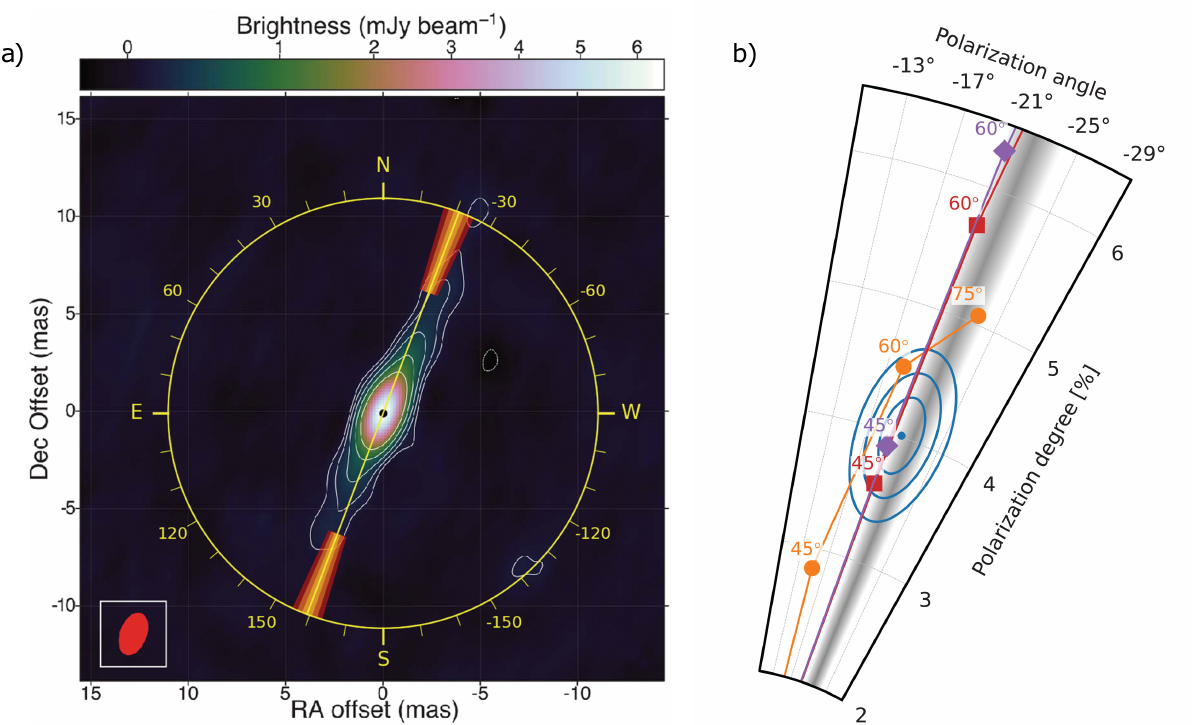}
\vspace*{-0.4cm}
\end{center}
\caption{Figures from Krawczynski et. al 2022 \cite{Krawczynski2022A}. (a) {\it IXPE} PA of Cygnus X-1 in the 2-8 keV energy band during the LHS observation overlaid on top of the radio jet imaged by the VLBA \cite{Miller-Jones2021}. Yellow, orange, and red bands give the PA at the 68\%, 95\% and 99.7\% confidence levels. (b) Comparison of the measured PD and PA (blue) with simulated polarization results for different corona configurations: sandwich coronae (orange) and hot inner flow--truncated disk corona with disk seed photons (red) or synchrotron seed photons (purple). Blue ellipses represent measurements to the 68\%, 95\% and 99.7\% confidence levels.) 
}\label{fig:cygx1}
\end{figure}

Cygnus X-1 is a persistently bright and highly variable $21.2 \pm 2.2 M_\odot$~\cite{Miller-Jones2021} black hole (although reverberation studies suggest a mass of
$16.5 \pm 5 M_\odot$~\cite{Mastroserio2019}) located $2.22$ kiloparsecs away. It exhibits a radio jet---an outflow of ionized matter organized in two diametrically opposing beams extending along the axis of rotation of the black hole---making it a microquasar. {\it IXPE} observed Cyg X-1 between May 15th and 21st of 2022 for a total new exposure of $\sim 242$ ksec in the LHS \cite{Krawczynski2022A}. The Cyg X-1 LHS observation revealed a polarization degree (PD) of $4.01 \pm 0.20$\% at a polarization angle (PA) of $-20.7^\circ \pm 1.4^\circ$. Figure~\ref{fig:cygx1}(a) shows that the measured PA and the radio jet of the source are almost perfectly aligned. The X-ray polarization in the {\it IXPE} energy band, which mainly comes from the inner disk, is perpendicular to the inner accretion disk plane according to our modelling.
The alignment of the PA and the radio jet in the plane of the sky indicates that the jet too is perpendicular to the inner accretion flow. This finding is consistent with the hypothesis that jets are launched from the inner regions of the system.

Different coronal geometries were tested using the {\tt kerrC} \cite{Krawczynski2012} and {\tt MONK} \cite{MONK} raytracing codes to replicate the observed high PD and correct PA alignment. We found that models in which the corona was vertically extended along the black hole spin axis (such as the cone-shaped and lamp-post coronae models) either predicted polarization degrees significantly below those observed and/or yielded polarization angles perpendicular to the jet axis. On the other hand, coronae extended laterally on the plane of the accretion disk predicted results more in accordance to the observation. The two geometries that were able to reproduce the data were the wedge-shaped/sandwich corona and the hot inner flow-truncated disk corona shown in Figure~\ref{fig:cygx1}(b). These models were able to explain the high, energy-dependent PD of the observation as well as produce an energy-independent polarization vector perpendicular to the accretion disk plane or parallel to the radio jet.

\subsection{Cygnus X-3}

\begin{figure}[t!]
\begin{center}
\includegraphics[width=15cm]{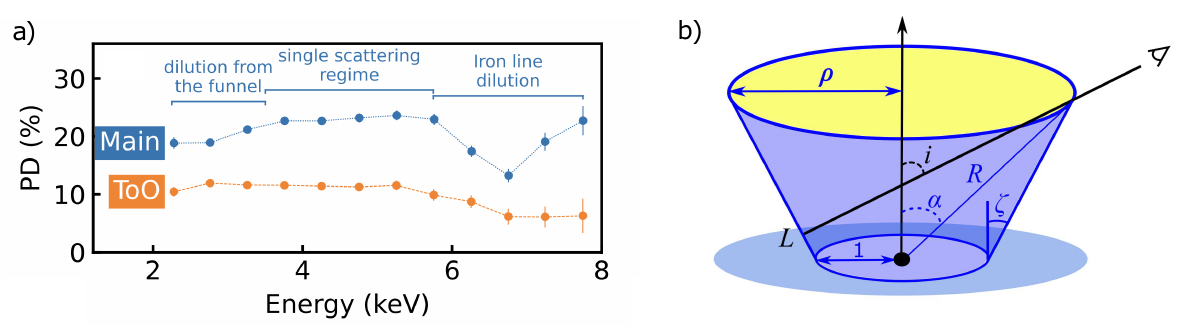}
\vspace*{-0.4cm}
\end{center}
\caption{Figures from Veledina et. al 2023 \cite{Veledina2023}. (a) {\it IXPE} PD vs energy of Cygnus X-3 in the 2-8 keV energy band during the LHS observation (Main, blue) and the intermediate, softer observation (ToO, orange). Error bars are given to 1$\sigma$.
(b) Schematic showing the proposed geometry of the collimating funnel region obscuring the primary source. An observer at inclination $i$ will see the reflection from the inner surface of the funnel until point $L$. $\zeta$ is the half-opening angle of the funnel, $R$ is the distance to the X-ray photosphere, and $\alpha$ is the angle at which the upper boundary of the funnel is seen from the primary X-ray source.}\label{fig:cygx3}
\end{figure}

Cygnus X-3 is a persistent black hole candidate at about 9.7 kiloparsecs from us \cite{Reid2023}. It is located in a binary with a Wolf-Rayet star and it is the brightest radio source among X-ray binaries \cite{McCollough1999, Belczynski2013}. 
From October 14th to October 19th and from October 31st to November 6th of 2022, {\it IXPE} observed Cyg X-3 for $\sim 538$ ksec and measured a PD of $20.6 \pm 0.3$\% at a PA of $90.1^\circ \pm 0.4^\circ$~\cite{Veledina2023}. This main observation occurred while the source was in the hard X-ray, radio quiescent state.
A second observational campaign for the source was later conducted from December 25th to December 29th, 2022 for $\sim 199$ ksec as the source transitioned to a softer state. 
For this Target of Opportunity (ToO) observation, {\it IXPE} measured a PD of $10.0 \pm 0.5$\% at a PA of $90.6^\circ \pm 1.2^\circ$ \cite{Veledina2023}. The PA in both observations was orthogonal to discrete radio ejections reported for the source suggesting that the polarization vector is approximately aligned with the plane of the accretion disk. Figure~\ref{fig:cygx3}(a) shows that the PDs for both observations and their negligible energy dependence. In fact, the PD only decreases at around $6.4$~keV where we expect the Fe K$\alpha$ emission line to depolarize the signal \cite{Marin2014}. The very high PD of the hard state observation and its orientation suggested that the observed emission was comprised mostly from reflected emission.

Correspondingly, Veledina et. al 2023 \cite{Veledina2023} posited that an optically thick medium shaped as a funnel must be collimating the radiation emitted by the accretion disk along its walls. Figure~\ref{fig:cygx3}(b) shows the proposed funnel region. The $\sim 10 \%$ change in polarization during the ToO observation when the source softened can then be explained as the outflow becoming more transparent. Given this geometry, the observed PD corresponds to a scattering angle of $\sim  38^{\circ}$ for which the model constrains the half-opening angle to $\alpha \lesssim 15^{\circ}$. Under these conditions, the apparent luminosity for an observer looking down at point $L$ in the funnel was estimated to be $L \gtrsim 5.5 x 10^{39}$ erg s$^{-1}$ which would make Cyg X-3 an obscured galactic ultraluminous X-ray source.

\subsection{LMC X-1}

\begin{figure}[t!]
\begin{center}
\includegraphics[width=15cm]{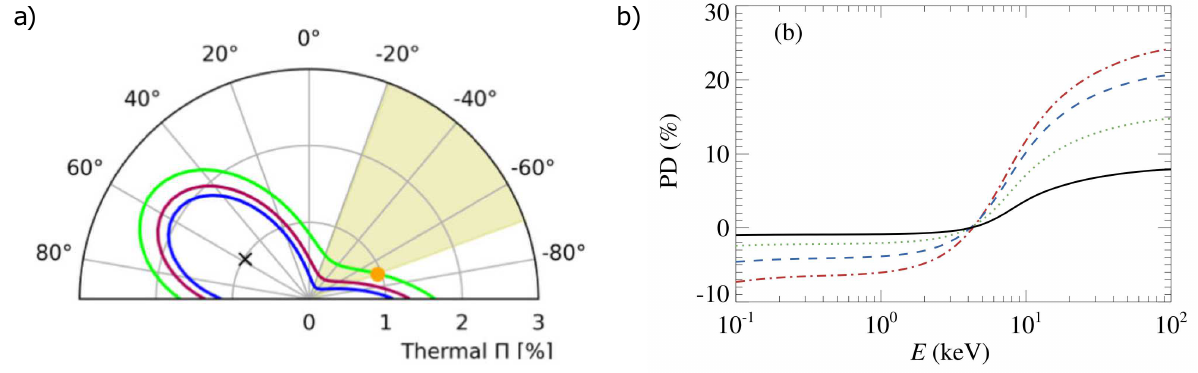}
\vspace*{-0.4cm}
\end{center}
\caption{Figures from Podgorn{\'y} et. al 2023 \cite{Podgorny2023}. (a) Contour plot of the LMC X-1 PD and PA of the thermal disk component when the PD of the coronal component is set to 0\% and their relative PAs are forced to be perpendicular. The blue, red, and grees line indicate 68\%, 90\% and 99\% confidence regions, respectively. The orange dot represents the 3$\sigma$ upper limits of the thermal emission polarization degree if we assume it is polarized in the direction of the projected accretion disk plane (yellow).
(b) PD predicted by the slab corona model of 10 keV electron temperature and $\tau_T= 1.26$ Thomson optical depth. Positive/Negative PD values correspond to a PA direction perpendicular/parallel to the disk.The black solid, green dotted, blue dashed, and red dot-dashed lines represent inclinations of $i=$ 30$^\circ$, 45$^\circ$, 60$^\circ$, and 75$^\circ$, respectively.
}\label{fig:lmcx1}
\end{figure}

LMC X-1 is an extragalactic $10.91 \pm 1.41 M_\odot~$~\cite{Orosz2009} black hole located 50 kiloparsecs away in the Large Magellanic Cloud \cite{Pietrzynski2013}. The source accretes matter from a $10.91 \pm 1.41 M_\odot$ early-type star making the system a high-mass X-ray binary \cite{White1985, Orosz2009}.
{\it IXPE} observed LMC X-1 between October 19th and October 28th, 2022 for a total of $\sim 562$ ksec \cite{Podgorny2023}. The observation took place while the source was in the HSS. The measured polarization was below the minimum detectable polarization of $1.1$\% at the 99\% confidence level for the 2-8 keV polarization so only an upper limit was imposed. The measured PD was $1.0 \pm 0.4$\% yielding a 3$\sigma$ upper limit of $2.2$\% while the PA suggested alignment with the projected ionization cone. 
Podgorn{\'y} et. al 2023 \cite{Podgorny2023} analyzed quasi-simultaneous NICER and {\it NuSTAR} data to decompose the total {\it IXPE} spectra into thermal disk and coronal emission contributions. In order to investigate how the thermal component was polarized, the polarization of the Comptonized component was set to 0\%, 4\%, and 10\% and the PAs of both components were forced to be perpendicular to each other. Figure~\ref{fig:lmcx1}(a) shows the contour plot for one of the configurations where the polarization of the coronal emission is 0\%. In this instance, the thermal emission PD upper limit at the 99\% confidence level is 2.2\% and 1.0\% if we require this component to be polarized in the direction of the accretion disk. When the PD of the coronal component is set to 4\% and 10\%, the PD of the thermal component if oriented in the plane of the disk reduces to 0.9\%.

Podgorn{\'y} et. al 2023 \cite{Podgorny2023} also employed a slab coronal geometry to understand the effect of the relative contributions of the thermal disk and coronal emissions to the total PD. Figure~\ref{fig:lmcx1}(b) shows that at $\sim 5$ keV, the change from negative to positive PD (which corresponds to a change of total polarization direction from parallel to perpendicular to the disk) occurs in the {\it IXPE} 2-8 keV band. As expected, competing components of parallel and perpendicular polarization tend to depolarize the radiation emitted from the innermost regions of the disk \cite[(see Fig. 4)]{Schnittman2009} and this could explain the overall low polarization signal from the source.

\subsection{4U 1630-47}

\begin{figure}[t!]
\begin{center}
\includegraphics[width=15cm]{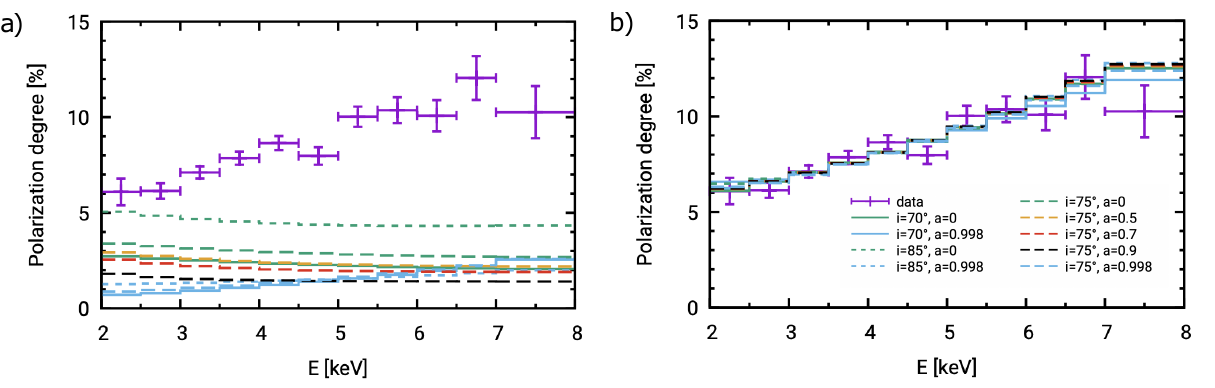}
\vspace*{-0.4cm}
\end{center}
\caption{Figure 8 from Ratheesh et. al 2023 \cite{Ratheesh2023}. 4U 1630-47 {\it IXPE} PD (purple) at the 68\% confidence level and model predictions. Solid, dashed, and dotted lines represent inclinations of 70$^\circ$, 75$^\circ$, and 85$^\circ$, respectively; green, yellow, red, black, and light blue colors represent spins of 0, 0.5, 0.7, 0.9 and 0.998, respectively. (a) PD predictions of a standard Novikov-Thorne thin disk with Chandrasekhar's pure electron-scattering atmosphere. (b) PD predictions for a thin disk with an outflowing, partially-ionized atmosphere of optical thickness $\tau = 7$ and velocity $v \sim 0.5 c$.
}\label{fig:4UHSS}
\end{figure}

4U 1630-47 is an X-ray transient source with outbursts every $\sim 600$ days \cite{Forman1976, Jones1976}. It is located about 4.7-11.5 kiloparsecs away \cite{Kalemci2018} and its parameters remain largely unconstrained due to extinction in the line of sight \cite{Parmar1986}.
The first {\it IXPE} observation of 4U 1630-47 was from August 23rd to September 2nd, 2022 for a total exposure time of $\sim 460$ ksec. {\it IXPE} caught the source in the HSS and measured a PD of $8.32 \pm 0.17$\% at a PA of $17.8^\circ \pm 0.6^\circ$ \cite{Ratheesh2023}. Spectral fitting of quasi-simultaneous NICER data revealed that the thermal contribution to the total emission was about 97\%. The PD linearly increased with energy from 6\% at 2~keV to 10\% at 8~keV while the PA showed no appreciable energy dependence. The measured PD was considerably higher
than the prediction for a pure electron-scattering atmosphere at the estimated $65^\circ$ inclination of the source---which should place the PD at around 3\% \cite{Chandrasekhar1960}. Figure~\ref{fig:4UHSS}(a) shows that the PD of the standard thin disk model under-predicts the observed PD.
Since the high rise of the PD with energy could not be explained by the standard thin disk model alone, Ratheesh et. al 2023 \cite{Ratheesh2023} propose a thin disk surrounded by a partially-ionized, outflowing atmosphere.

In this model, the partial ionization of the disk photosphere increases the PD as absorption processes lead to an increased number photons travelling perpendicular to the disk plane \cite{Taverna2021}. Additionally, assuming a vertical outflow velocity requires a higher local emission angle to reach the same observer at a fixed inclination due to aberration effects---also resulting in a higher PD. Figure~\ref{fig:4UHSS}(b) shows the PDs predicted for this model for partially-ionized atmosphere of optical thickness $\tau = 7$ outflowing at a $v \sim 0.5 c$. This model predicts a PA parallel to the disk, although the lack of radio information on the source's jet makes it currently impossible to verify this result.

\begin{figure}[t!]
\begin{center}
\includegraphics[width=12cm]{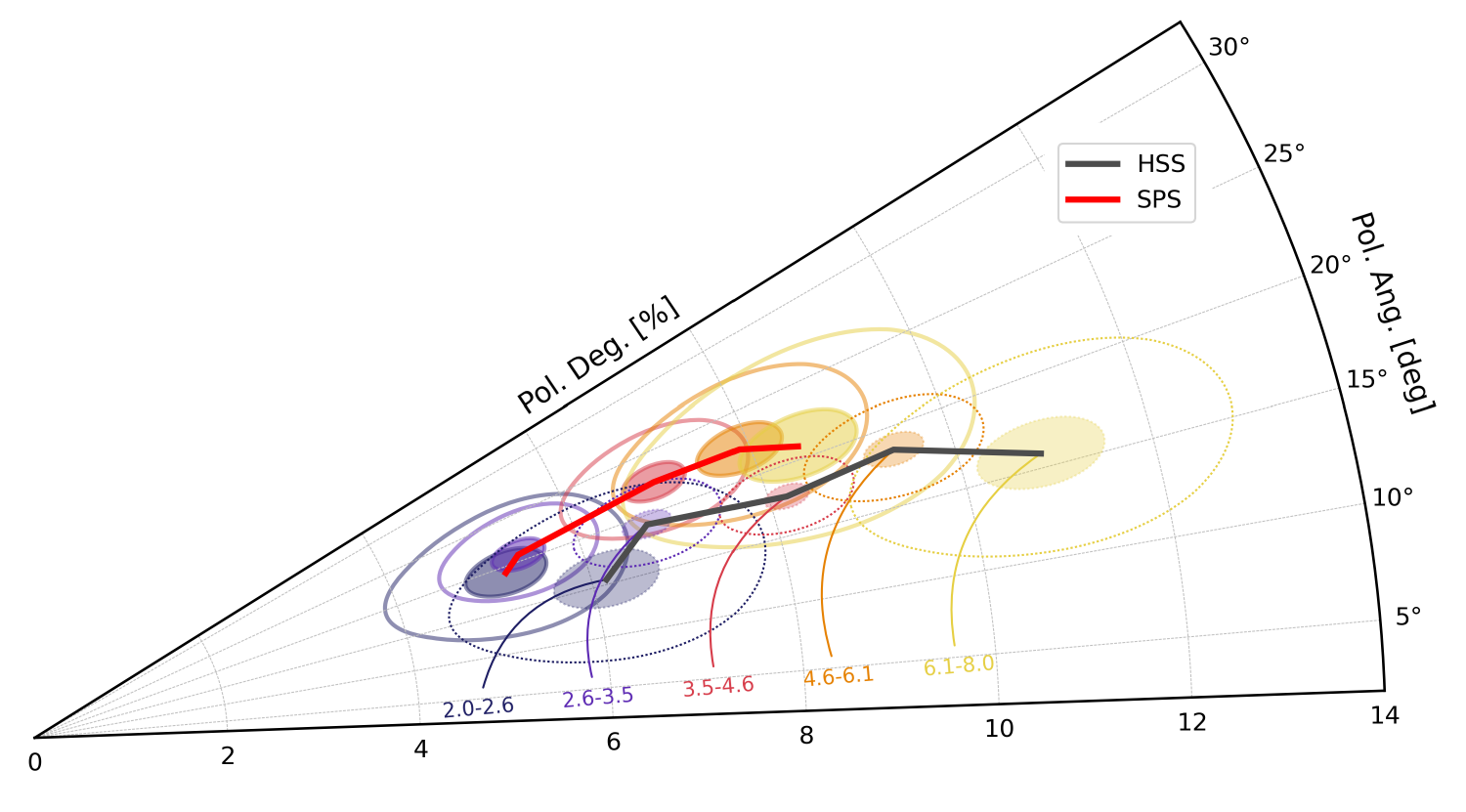}
\vspace*{-0.4cm}
\end{center}
\caption{Figure 3 of Rodriguez Cavero et. al 2023 \cite{RodriguezCavero2023}. PD and PA of 4U 1630-47 in the HSS (black) and SPL (red) states binned in 5 logarithmic energy bins. The shaded and unshaded ellipses are confidence contours a the 68.3\% and 99.7\% confidence levels.
}\label{fig:4USPL}\vspace*{-0.4cm}
\end{figure}

A second {\it IXPE} observation of 4U 1630-47 occurred after MAXI reported a significant increase in flux hinting that the source was in the SPL state. {\it IXPE} observed the source from March 10th to March 14th, 2023 for $\sim 140$ ksec and measured a PD of $6.8 \pm 0.2$\% at a PA of $21.3^\circ \pm 0.9^\circ$ \cite{RodriguezCavero2023}. The thermal contributions of the source reported in this state were between 8\% and 54\% depending on the spectral models used to fit NICER and {\it NuSTAR} data acquired during the {\it IXPE} observation, while the rest of the emission is attributed to a power-law component. The polarization results for both {\it IXPE} observations of 4U 1630-47 are shown in Figure~\ref{fig:4USPL}. The small 1.5\% change in the PD and the constant PA (within $3 \sigma$) during the state transition from the HSS to the SPL state were unexpected. The authors attribute these results to an instantaneous increase in electron temperature leading to a change from Thomson scattering processes in the partially-ionized accretion disk atmosphere to inverse Compton scattering that leads to the observed power-law component.

\section{Discussion and Conclusions}

The first year of {\it IXPE} stellar-mass black hole observations demonstrated that polarization measurements are necessary to probe the complex environment surrounding these astrophysical sources. In 2022, {\it IXPE} observed four sources: Cyg X-1, Cyg X-3, LMC X-1, and 4U 1630-47. While the first two sources were observed in the harder states of accretion, their polarization properties show vastly different emission geometries. In the case of Cyg X-1, we saw predominantly coronal emission and an energy-dependent PD that allowed us to constrain the corona geometry to laterally extended in the plane of the disk \cite{Krawczynski2022A}. For Cyg X-3, polarization results showed that most of the emission came from reflection. Cyg X-3 high PD measurement suggested that the disk emission is collimated along the walls of an optically thick funnel region and that the black hole candidate might be an obscured ultra-luminous X-ray source \cite{Veledina2023}.

Although we were not able to achieve a statistically significant polarization measurement of LMC X-1, we obtained an upper limit of the total polarization signal \cite{Podgorny2023}. LMC X-1 spends most of its time in the HSS and the derived upper limits for the dominant thermal emission were in line with those predicted by Chandrasekhar for a semi-infinite scattering atmosphere \cite{Chandrasekhar1960}. However, like Cyg X-1, the PA of LMC X-1 seems to be aligned to the projected ionization cone structure. This is puzzling since the thermal emission is 94\% of the total emission and is expected to be polarized perpendicular to the ionization cone.
4U 1630-47 was also measured in the HSS and its PD was considerably larger than those estimates by the standard thin disk model \cite{Ratheesh2023}. In this case, the polarization 4U 1630-47 required a partially-ionized, optically thick outflowing atmosphere to achieve that higher PD and yielded a PA parallel to the disk. The source was later observed again in the SPL state \cite{RodriguezCavero2023} where there was a small decrease in the PD and a constant PA even though a power-law component should introduce a greater PD at a perpendicular direction to that of direct emission.

\pagebreak

{\it IXPE} observations of BHXRBs during 2022 revealed (for the most part) high PDs and persistent PAs. The proposed outflow geometries that deviate from the standard thin disk-corona configuration underscore the need for a more comprehensive investigation of the plasma surrounding the inner accretion flows. Changes in the winds of the system, evolution of the corona during state transitions, and slim disk models are currently being considered.
Future {\it IXPE} observation of stellar-mass black holes will allow us to discern between models of accretion.

\acknowledgments{
The Imaging X-ray Polarimetry Explorer (IXPE) is a joint US and Italian mission.  The US contribution is supported by the National Aeronautics and Space Administration (NASA) and led and managed by its Marshall Space Flight Center (MSFC), with industry partner Ball Aerospace (contract NNM15AA18C).  The Italian contribution is supported by the Italian Space Agency (Agenzia Spaziale Italiana, ASI) through contract ASI-OHBI-2022-13-I.0, agreements ASI-INAF-2022-19-HH.0 and ASI-INFN-2017.13-H0, and its Space Science Data Center (SSDC) with agreements ASI-INAF-2022-14-HH.0 and ASI-INFN 2021-43-HH.0, and by the Istituto Nazionale di Astrofisica (INAF) and the Istituto Nazionale di Fisica Nucleare (INFN) in Italy.  This research used data products provided by the IXPE Team (MSFC, SSDC, INAF, and INFN) and distributed with additional software tools by the High-Energy Astrophysics Science Archive Research Center (HEASARC), at NASA Goddard Space Flight Center (GSFC).}



\bibliographystyle{JHEP}
\bibliography{refs}

\providecommand{\href}[2]{#2}\begingroup\raggedright\begin{thebibliography}{10}

\bibitem{Shakura-Sunyaev1973}
N.I.~{Shakura} and R.A.~{Sunyaev}, \emph{{Black holes in binary systems. Observational appearance.}}, {\emph{Astronomy \& Astrophysics} {\bfseries 24} (1973) 337}.

\bibitem{Done2007}
C.~{Done}, M.~{Gierli{\'n}ski} and A.~{Kubota}, \emph{{Modelling the behaviour of accretion flows in X-ray binaries. Everything you always wanted to know about accretion but were afraid to ask}}, \href{https://doi.org/10.1007/s00159-007-0006-1}{\emph{Astronomy \& Astrophysics Review} {\bfseries 15} (2007) 1} [\href{https://arxiv.org/abs/0708.0148}{{\ttfamily 0708.0148}}].

\bibitem{McClintock2006}
J.E.~{McClintock} and R.A.~{Remillard}, \emph{{Black hole binaries}},  in \emph{Compact stellar X-ray sources}, vol.~39, pp.~157--213, Cambridge University Press (2006), \href{https://doi.org/10.48550/arXiv.astro-ph/0306213}{DOI}.

\bibitem{Schnittman2010}
J.D.~{Schnittman} and J.H.~{Krolik}, \emph{{X-ray Polarization from Accreting Black Holes: Coronal Emission}}, \href{https://doi.org/10.1088/0004-637X/712/2/908}{\emph{The Astrophysical Journal} {\bfseries 712} (2010) 908} [\href{https://arxiv.org/abs/0912.0907}{{\ttfamily 0912.0907}}].

\bibitem{Zhang2022}
W.~{Zhang} et~al., \emph{{Investigating the X-ray polarization of lamp-post coronae in BHXRBs}}, \href{https://doi.org/10.1093/mnras/stac1937}{\emph{Monthly Notices of the Royal Astronomical Society} {\bfseries 515} (2022) 2882} [\href{https://arxiv.org/abs/2207.03228}{{\ttfamily 2207.03228}}].

\bibitem{Gilfanov2010}
M.~{Gilfanov}, \emph{{X-Ray Emission from Black-Hole Binaries}},  in \emph{Lecture Notes in Physics, Berlin Springer Verlag}, T.~{Belloni}, ed., vol.~794, p.~17 (2010), \href{https://doi.org/10.1007/978-3-540-76937-8_2}{DOI}.

\bibitem{Fender2004}
R.P.~{Fender}, T.M.~{Belloni} and E.~{Gallo}, \emph{{Towards a unified model for black hole X-ray binary jets}}, \href{https://doi.org/10.1111/j.1365-2966.2004.08384.x}{\emph{Monthly Notices of the Royal Astronomical Society} {\bfseries 355} (2004) 1105} [\href{https://arxiv.org/abs/astro-ph/0409360}{{\ttfamily astro-ph/0409360}}].

\bibitem{Mitsuda1984}
K.~{Mitsuda}, H.~{Inoue}, K.~{Koyama}, K.~{Makishima}, M.~{Matsuoka}, Y.~{Ogawara} et~al., \emph{{Energy spectra of low-mass binary X-ray sources observed from Tenma.}}, {\emph{Publications of the Astronomical Society of Japan} {\bfseries 36} (1984) 741}.

\bibitem{Tomsick2014A}
J.A.~{Tomsick} et~al., \emph{{The Reflection Component from Cygnus X-1 in the Soft State Measured by NuSTAR and Suzaku}}, \href{https://doi.org/10.1088/0004-637X/780/1/78}{\emph{The Astrophysical Journal} {\bfseries 780} (2014) 78} [\href{https://arxiv.org/abs/1310.3830}{{\ttfamily 1310.3830}}].

\bibitem{Szostek2008}
A.~{Szostek}, A.A.~{Zdziarski} and M.L.~{McCollough}, \emph{{A classification of the X-ray and radio states of Cyg X-3 and their long-term correlations}}, \href{https://doi.org/10.1111/j.1365-2966.2008.13479.x}{\emph{Monthly Notices of the Royal Astronomical Society} {\bfseries 388} (2008) 1001} [\href{https://arxiv.org/abs/0803.2217}{{\ttfamily 0803.2217}}].

\bibitem{Zdziarski2004}
A.A.~{Zdziarski} and M.~{Gierli{\'n}ski}, \emph{{Radiative Processes, Spectral States and Variability of Black-Hole Binaries}}, \href{https://doi.org/10.1143/PTPS.155.99}{\emph{Progress of Theoretical Physics Supplement} {\bfseries 155} (2004) 99} [\href{https://arxiv.org/abs/astro-ph/0403683}{{\ttfamily astro-ph/0403683}}].

\bibitem{Remillard2006}
R.A.~{Remillard} and J.E.~{McClintock}, \emph{{X-Ray Properties of Black-Hole Binaries}}, \href{https://doi.org/10.1146/annurev.astro.44.051905.092532}{\emph{Annual Review of Astronomy and Astrophysics} {\bfseries 44} (2006) 49} [\href{https://arxiv.org/abs/astro-ph/0606352}{{\ttfamily astro-ph/0606352}}].

\bibitem{Schnittman2009}
J.D.~{Schnittman} and J.H.~{Krolik}, \emph{{X-ray Polarization from Accreting Black Holes: The Thermal State}}, \href{https://doi.org/10.1088/0004-637X/701/2/1175}{\emph{The Astrophysical Journal} {\bfseries 701} (2009) 1175} [\href{https://arxiv.org/abs/0902.3982}{{\ttfamily 0902.3982}}].

\bibitem{Krawczynski2012}
H.~{Krawczynski}, \emph{{Tests of General Relativity in the Strong-gravity Regime Based on X-Ray Spectropolarimetric Observations of Black Holes in X-Ray Binaries}}, \href{https://doi.org/10.1088/0004-637X/754/2/133}{\emph{The Astrophysical Journal} {\bfseries 754} (2012) 133} [\href{https://arxiv.org/abs/1205.7063}{{\ttfamily 1205.7063}}].

\bibitem{MONK}
W.~{Zhang}, M.~{Dov{\v{c}}iak} and M.~{Bursa}, \emph{{Constraining the Size of the Corona with Fully Relativistic Calculations of Spectra of Extended Coronae. I. The Monte Carlo Radiative Transfer Code}}, \href{https://doi.org/10.3847/1538-4357/ab1261}{\emph{The Astrophysical Journal} {\bfseries 875} (2019) 148} [\href{https://arxiv.org/abs/1903.09241}{{\ttfamily 1903.09241}}].

\bibitem{Chandrasekhar1960}
S.~{Chandrasekhar}, \emph{{Radiative transfer}}, Dover Publications (1960).

\bibitem{Dovciak2008}
M.~{Dov{\v{c}}iak}, F.~{Muleri}, R.W.~{Goosmann}, V.~{Karas} and G.~{Matt}, \emph{{Thermal disc emission from a rotating black hole: X-ray polarization signatures}}, \href{https://doi.org/10.1111/j.1365-2966.2008.13872.x}{\emph{Monthly Notices of the Royal Astronomical Society} {\bfseries 391} (2008) 32} [\href{https://arxiv.org/abs/0809.0418}{{\ttfamily 0809.0418}}].

\bibitem{Li2009}
L.-X.~{Li}, R.~{Narayan} and J.E.~{McClintock}, \emph{{Inferring the Inclination of a Black Hole Accretion Disk from Observations of its Polarized Continuum Radiation}}, \href{https://doi.org/10.1088/0004-637X/691/1/847}{\emph{The Astrophysical Journal} {\bfseries 691} (2009) 847} [\href{https://arxiv.org/abs/0809.0866}{{\ttfamily 0809.0866}}].

\bibitem{Weisskopf2022}
M.C.~{Weisskopf}, P.~{Soffitta}, L.~{Baldini}, B.D.~{Ramsey}, S.L.~{O'Dell}, R.W.~{Romani} et~al., \emph{{The Imaging X-Ray Polarimetry Explorer (IXPE): Pre-Launch}}, \href{https://doi.org/10.1117/1.JATIS.8.2.026002}{\emph{Journal of Astronomical Telescopes, Instruments, and Systems} {\bfseries 8} (2022) 026002} [\href{https://arxiv.org/abs/2112.01269}{{\ttfamily 2112.01269}}].

\bibitem{OSO-8}
H.L.~{Kestenbaum}, G.G.~{Cohen}, K.S.~{Long}, R.~{Novick}, E.H.~{Silver}, M.C.~{Weisskopf} et~al., \emph{{The graphite crystal X-ray spectrometer on OSO-8.}}, \href{https://doi.org/10.1086/154889}{\emph{The Astrophysical Journal} {\bfseries 210} (1976) 805}.

\bibitem{Soffita2020}
P.~{Soffitta}, P.~{Attin{\`a}}, L.~{Baldini}, M.~{Barbanera}, W.H.~{Baumgartner}, R.~{Bellazzini} et~al., \emph{{The Imaging X-ray Polarimetry Explorer (IXPE): technical overview III}},  in \emph{SPIE}, vol.~11444 of \emph{Society of Photo-Optical Instrumentation Engineers (SPIE) Conference Series}, Dec., 2020, \href{https://doi.org/10.1117/12.2567001}{DOI}.

\bibitem{DiMarco2022}
A.~{Di Marco}, S.~{Fabiani}, F.~{La Monaca}, F.~{Muleri}, J.~{Rankin}, P.~{Soffitta} et~al., \emph{{Calibration of the IXPE Focal Plane X-Ray Polarimeters to Polarized Radiation}}, \href{https://doi.org/10.3847/1538-3881/ac7719}{\emph{The Astrophysical Journal} {\bfseries 164} (2022) 103} [\href{https://arxiv.org/abs/2206.07582}{{\ttfamily 2206.07582}}].

\bibitem{nicer}
K.C.~{Gendreau}, Z.~{Arzoumanian} and T.~{Okajima}, \emph{{The Neutron star Interior Composition ExploreR (NICER): an Explorer mission of opportunity for soft x-ray timing spectroscopy}},  in \emph{Space Telescopes and Instrumentation 2012: Ultraviolet to Gamma Ray}, T.~{Takahashi}, S.S.~{Murray} and J.-W.A.~{den Herder}, eds., vol.~8443 of \emph{Society of Photo-Optical Instrumentation Engineers (SPIE) Conference Series}, p.~844313, Sept., 2012, \href{https://doi.org/10.1117/12.926396}{DOI}.

\bibitem{nustar}
F.A.~{Harrison}, W.W.~{Craig}, F.E.~{Christensen}, C.J.~{Hailey}, W.W.~{Zhang}, S.E.~{Boggs} et~al., \emph{{The Nuclear Spectroscopic Telescope Array (NuSTAR) High-energy X-Ray Mission}}, \href{https://doi.org/10.1088/0004-637X/770/2/103}{\emph{The Astrophysical Journal} {\bfseries 770} (2013) 103} [\href{https://arxiv.org/abs/1301.7307}{{\ttfamily 1301.7307}}].

\bibitem{Miller-Jones2021}
J.C.A.~{Miller-Jones}, A.~{Bahramian}, J.A.~{Orosz}, I.~{Mandel}, L.~{Gou}, T.J.~{Maccarone} et~al., \emph{{Cygnus X-1 contains a 21-solar mass black hole{\textemdash}Implications for massive star winds}}, \href{https://doi.org/10.1126/science.abb3363}{\emph{Science} {\bfseries 371} (2021) 1046} [\href{https://arxiv.org/abs/2102.09091}{{\ttfamily 2102.09091}}].

\bibitem{Mastroserio2019}
G.~{Mastroserio}, A.~{Ingram} and M.~{van der Klis}, \emph{{An X-ray reverberation mass measurement of Cygnus X-1}}, \href{https://doi.org/10.1093/mnras/stz1727}{\emph{Monthly Notices of the Royal Astronomical Society} {\bfseries 488} (2019) 348} [\href{https://arxiv.org/abs/1906.08266}{{\ttfamily 1906.08266}}].

\bibitem{Krawczynski2022A}
H.~{Krawczynski}, F.~{Muleri}, M.~{Dov{\v{c}}iak}, A.~{Veledina}, N.~{Rodriguez Cavero}, J.~{Svoboda} et~al., \emph{{Polarized x-rays constrain the disk-jet geometry in the black hole x-ray binary Cygnus X-1}}, \href{https://doi.org/10.1126/science.add5399}{\emph{Science} {\bfseries 378} (2022) 650} [\href{https://arxiv.org/abs/2206.09972}{{\ttfamily 2206.09972}}].

\bibitem{Reid2023}
M.J.~{Reid} and J.C.A.~{Miller-Jones}, \emph{{On the Distances to the X-Ray Binaries Cygnus X-3 and GRS 1915+105}}, \href{https://doi.org/10.3847/1538-4357/acfe0c}{\emph{The Astrophysical Journal} {\bfseries 959} (2023) 85} [\href{https://arxiv.org/abs/2309.15027}{{\ttfamily 2309.15027}}].

\bibitem{McCollough1999}
M.L.~{McCollough}, C.R.~{Robinson}, S.N.~{Zhang}, B.A.~{Harmon}, R.M.~{Hjellming}, E.B.~{Waltman} et~al., \emph{{Discovery of Correlated Behavior between the Hard X-Ray and the Radio Bands in Cygnus X-3}}, \href{https://doi.org/10.1086/307241}{\emph{The Astrophysical Journal} {\bfseries 517} (1999) 951} [\href{https://arxiv.org/abs/astro-ph/9810212}{{\ttfamily astro-ph/9810212}}].

\bibitem{Belczynski2013}
K.~{Belczynski}, T.~{Bulik}, I.~{Mandel}, B.S.~{Sathyaprakash}, A.A.~{Zdziarski} and J.~{Miko{\l}ajewska}, \emph{{Cyg X-3: A Galactic Double Black Hole or Black-hole-Neutron-star Progenitor}}, \href{https://doi.org/10.1088/0004-637X/764/1/96}{\emph{The Astrophysical Journal} {\bfseries 764} (2013) 96} [\href{https://arxiv.org/abs/1209.2658}{{\ttfamily 1209.2658}}].

\bibitem{Veledina2023}
A.~{Veledina}, F.~{Muleri}, J.~{Poutanen}, J.~{Podgorn{\'y}}, M.~{Dov{\v{c}}iak}, F.~{Capitanio} et~al., \emph{{Astronomical puzzle Cyg X-3 is a hidden Galactic ultraluminous X-ray source}}, \href{https://doi.org/10.48550/arXiv.2303.01174}{\emph{arXiv e-prints} (2023) arXiv:2303.01174} [\href{https://arxiv.org/abs/2303.01174}{{\ttfamily 2303.01174}}].

\bibitem{Marin2014}
F.~{Marin} and F.~{Tamborra}, \emph{{Probing the origin of the iron K{\ensuremath{\alpha}} line around stellar and supermassive black holes using X-ray polarimetry}}, \href{https://doi.org/10.1016/j.asr.2013.09.003}{\emph{Advances in Space Research} {\bfseries 54} (2014) 1458} [\href{https://arxiv.org/abs/1309.1684}{{\ttfamily 1309.1684}}].

\bibitem{Orosz2009}
J.A.~{Orosz}, D.~{Steeghs}, J.E.~{McClintock}, M.A.P.~{Torres}, I.~{Bochkov}, L.~{Gou} et~al., \emph{{A New Dynamical Model for the Black Hole Binary LMC X-1}}, \href{https://doi.org/10.1088/0004-637X/697/1/573}{\emph{The Astrophysical Journal} {\bfseries 697} (2009) 573} [\href{https://arxiv.org/abs/0810.3447}{{\ttfamily 0810.3447}}].

\bibitem{Pietrzynski2013}
G.~{Pietrzy{\'n}ski}, D.~{Graczyk}, W.~{Gieren}, I.B.~{Thompson}, B.~{Pilecki}, A.~{Udalski} et~al., \emph{{An eclipsing-binary distance to the Large Magellanic Cloud accurate to two per cent}}, \href{https://doi.org/10.1038/nature11878}{\emph{Nature} {\bfseries 495} (2013) 76} [\href{https://arxiv.org/abs/1303.2063}{{\ttfamily 1303.2063}}].

\bibitem{White1985}
N.E.~{White} and K.O.~{Mason}, \emph{{The Structure of Low-Mass X-Ray Binaries}}, \href{https://doi.org/10.1007/BF00212883}{\emph{Space Science Reviews} {\bfseries 40} (1985) 167}.

\bibitem{Podgorny2023}
J.~{Podgorn{\'y}}, L.~{Marra}, F.~{Muleri}, N.~{Rodriguez Cavero}, A.~{Ratheesh}, M.~{Dov{\v{c}}iak} et~al., \emph{{The first X-ray polarimetric observation of the black hole binary LMC X-1}}, \href{https://doi.org/10.1093/mnras/stad3103}{\emph{Monthly Notices of the Royal Astronomical Society} (2023) }.

\bibitem{Forman1976}
W.~{Forman}, C.~{Jones} and H.~{Tananbaum}, \emph{{Uhuru observations of a transient X-ray source associated with the globular cluster NGC 6440.}}, \href{https://doi.org/10.1086/182170}{\emph{The Astrophysical Journal Letters} {\bfseries 207} (1976) L25}.

\bibitem{Jones1976}
C.~{Jones}, W.~{Forman}, H.~{Tananbaum} and M.J.L.~{Turner}, \emph{{Uhuru and Ariel V observations of 3U 1630-47: a recurrent transient X-ray source.}}, \href{https://doi.org/10.1086/182291}{\emph{The Astrophysical Journal Letters} {\bfseries 210} (1976) L9}.

\bibitem{Kalemci2018}
E.~{Kalemci}, T.J.~{Maccarone} and J.A.~{Tomsick}, \emph{{A Dust-scattering Halo of 4U 1630-47 Observed with Chandra and Swift: New Constraints on the Source Distance}}, \href{https://doi.org/10.3847/1538-4357/aabcd3}{\emph{The Astrophysical Journal} {\bfseries 859} (2018) 88} [\href{https://arxiv.org/abs/1804.02909}{{\ttfamily 1804.02909}}].

\bibitem{Parmar1986}
A.N.~{Parmar}, L.~{Stella} and N.E.~{White}, \emph{{The Evolution of the 1984 Outburst of the Transient X-Ray Source 4U 1630-47}}, \href{https://doi.org/10.1086/164204}{\emph{The Astrophysical Journal} {\bfseries 304} (1986) 664}.

\bibitem{Ratheesh2023}
A.~{Ratheesh}, M.~{Dov{\v{c}}iak}, H.~{Krawczynski}, J.~{Podgorn{\'y}}, L.~{Marra}, A.~{Veledina} et~al., \emph{{The high polarisation of the X-rays from the Black Hole X-ray Binary 4U 1630-47 challenges standard thin accretion disc scenario}}, \href{https://doi.org/10.48550/arXiv.2304.12752}{\emph{arXiv e-prints} (2023) arXiv:2304.12752} [\href{https://arxiv.org/abs/2304.12752}{{\ttfamily 2304.12752}}].

\bibitem{Taverna2021}
R.~{Taverna}, L.~{Marra}, S.~{Bianchi}, M.~{Dov{\v{c}}iak}, R.~{Goosmann}, F.~{Marin} et~al., \emph{{Spectral and polarization properties of black hole accretion disc emission: including absorption effects}}, \href{https://doi.org/10.1093/mnras/staa3859}{\emph{Monthly Notices of the Royal Astronomical Society} {\bfseries 501} (2021) 3393} [\href{https://arxiv.org/abs/2012.06504}{{\ttfamily 2012.06504}}].

\bibitem{RodriguezCavero2023}
N.~{Rodriguez Cavero}, L.~{Marra}, H.~{Krawczynski}, M.~{Dov{\v{c}}iak}, S.~{Bianchi}, J.F.~{Steiner} et~al., \emph{{The First X-Ray Polarization Observation of the Black Hole X-Ray Binary 4U 1630-47 in the Steep Power-law State}}, \href{https://doi.org/10.3847/2041-8213/acfd2c}{\emph{The Astrophysical Journal Letters} {\bfseries 958} (2023) L8} [\href{https://arxiv.org/abs/2305.10630}{{\ttfamily 2305.10630}}].

\end{thebibliography}\endgroup


\end{document}